\newcommand*{\ATLASLATEXPATH}{latex/} 
\documentclass[cernpreprint,texlive=2011,txfonts,UKenglish]{latex/atlasdoc} 
\pdfoutput=1

\usepackage{\ATLASLATEXPATH atlaspackage}
\usepackage{lscape}
\usepackage{\ATLASLATEXPATH atlasbiblatex}

\usepackage{\ATLASLATEXPATH atlascontribute}

\usepackage{\ATLASLATEXPATH atlasphysics}
\usepackage{\ATLASLATEXPATH atlasprocess}

\graphicspath{{logos/}{figures/}}

\AtlasTitle{Measurement of \Wboson and \Zboson-boson production cross sections in $pp$ collisions at $\sqrt{s}=13\TeV$ with the ATLAS detector}

\author{The ATLAS Collaboration}

\AtlasJournal{PLB}
\AtlasRefCode{atlas-STDM-2015-03}

\AtlasNote{ATL-COM-PHYS-2016-026}

\PreprintIdNumber{CERN-EP-2016-069}

\AtlasAbstract{
Measurements of the $\Wboson \to \ell^{\pm} \nu$ and \Ztoll\ production cross sections
(where $\ell^{\pm}=e^{\pm},\mu^{\pm}$) in proton--proton collisions at $\sqrt{s}=\SI{13}{\TeV}$ are presented 
using data recorded by the ATLAS experiment at the Large Hadron Collider, corresponding to a total 
integrated luminosity of \SI{81}{\per\pb}.
The total inclusive \Wboson-boson production cross sections times the single-lepton-flavour branching ratios 
are $\stotWp=$~$\AveTotCentralWp \pm \AveTotStatWp~\text{(stat)} \pm \AveTotSystWp~\text{(sys)} \pm \AveTotLumiWp~\text{(lumi)}$~nb
and $\stotWm=$~$\AveTotCentralWm \pm \AveTotStatWm~\text{(stat)} \pm \AveTotSystWm~\text{(sys)} \pm \AveTotLumiWm~\text{(lumi)}$~nb for \Wplus and \Wminus, 
respectively. The total inclusive \Zboson-boson production cross section times leptonic branching ratio, 
within the invariant mass window $66 < m_{\ell\ell} < \SI{116}{\GeV}$,  is $\sigma^\text{tot}_Z $~=~$\AveTotCentralZ \pm \AveTotStatZ~\text{(stat)} \pm \AveTotSystZ~\text{(sys)} \pm \AveTotLumiZ~\text{(lumi)}$~nb.
The \Wplus, \Wminus, and \Zboson-boson production cross sections and cross-section ratios within a fiducial region defined by the detector acceptance are also measured. 
The cross-section ratios benefit from significant cancellation of experimental uncertainties, resulting in 
$\sfidWp/\sfidWm= \WpWmcombVal \pm \WpWmcombSt~\text{(stat)} \pm \WpWmcombSy~\text{(sys)}$ and 
$ \sfidWpm/\sfidZ = \WZcombVal \pm \WZcombSt~\text{(stat)} \pm \WZcombSy~\text{(sys)}$.
Theoretical predictions, based on calculations accurate to next-to-next-to-leading order for quantum chromodynamics and to next-to-leading order for electroweak processes and which employ different parton distribution function sets, are compared to these measurements. 
}

\AtlasCoverSupportingNote{Support note of the measurement of the $W$ and $Z$ cross section at $\sqrt{}s=\SI{13}\TeV$}{https://cds.cern.ch/record/2060737} 
\AtlasCoverCommentsDeadline{2nd June 2016}

\AtlasCoverAnalysisTeam{Ludovica~Aperio~Bella, Lidia~Dell’Asta, Lorenzo~Bellagamba, Kevin~Black, Ulrike~Blumenschein, John~Butler, Stefano~Camarda, Valentina~Maria~Cairo, Chiara~Debenedetti, Marisilvia~Donadelli, Laura~Fabbri, Benedetto~Giacobbe, Alexandre~Glazov, Louis~Helary, Oleh~Kivernyk, Uta~Klein, Alexander~Law, Zhijun~Liang, Marco~Aurelio~Lisboa~Leite, Mykhailo~Lisovyi, Kristin~Lohwasser, Jason~Nielsen, Konstantinos~Nikolopoulos, Rhys~Edward~Owen, Hayk~Pirumov, Voica~Radescu, Eram~Rizvi, Kristof~Schmieden, Federico~Sforza, Alexander~Sherman, Simao~Paulo~Silva, Philip~Sommer, Artur~Trofymov, Manuella~Vincter, Camilla~Vittori, Efe~Yigitbasi, Nataliia~Zakharchuk}

\AtlasCoverEdBoardMember{Corrinne Mills~(chair), Jahred Adelman, Evelin Meoni}

\AtlasCoverEgroupEditors{atlas-STDM-2015-03-editors@cern.ch}

\AtlasCoverEgroupEdBoard{atlas-STDM-2015-03-editorial-board@cern.ch}

\hypersetup{pdftitle={ATLAS draft},pdfauthor={The ATLAS Collaboration}}

\newcommand{\sigfidZenolabel}{\ensuremath{  0.781 \pm    0.004  \pm    0.008  \pm   0.016  }} 

\newcommand{\sigZenolabel}{\ensuremath{ 1.987 \pm   0.011  \pm  0.041  \pm   0.042  }}
\newcommand{\CZe}{ 0.552}
 
\newcommand{\CZeSystUpErr}{+0.006}
\newcommand{\CZeSystDownErr}{-0.005} 
\newcommand{\ntotZe}{     35,009}

\newcommand{\ntotsignalZe}{$  34865 \pm    187 \pm     7 \pm 3 $}

\newcommand{\sigfidZmunolabel}{\ensuremath{  0.777 \pm    0.004  \pm    0.008  \pm   0.016  }} 

\newcommand{\sigZmunolabel}{\ensuremath{ 1.977 \pm    0.009  \pm   0.041  \pm   0.042  }}
\newcommand{\CZmu}{ 0.711}

\newcommand{\CZmuSyst}{0.008} 
\newcommand{\ntotZmu}{     44,898}

\newcommand{\ntotsignalZmu}{$  44706 \pm    212 \pm  9 \pm 4 $}

\newcommand{\AveCentralZ}{   0.779}
\newcommand{\AveStatZ}{      0.003}
\newcommand{\AveSystZ}{      0.006}
\newcommand{\AveLumiZ}{      0.016}
\newcommand{\AveTotCentralZ}{1.981}
\newcommand{\AveTotStatZ}{   0.007}
\newcommand{\AveTotSystZ}{   0.038}
\newcommand{\AveTotLumiZ}{   0.042}
\newcommand{\AveCentralWm}{  3.50}
\newcommand{\AveStatWm}{     0.01}
\newcommand{\AveSystWm}{     0.07}
\newcommand{\AveLumiWm}{     0.07}
\newcommand{\AveTotCentralWm}{8.79}
\newcommand{\AveTotStatWm}{   0.02}
\newcommand{\AveTotSystWm}{   0.24}
\newcommand{\AveTotLumiWm}{   0.18}
\newcommand{\AveCentralWp}{   4.53}
\newcommand{\AveStatWp}{      0.01}
\newcommand{\AveSystWp}{      0.09}
\newcommand{\AveLumiWp}{      0.10}
\newcommand{\AveTotCentralWp}{11.83}
\newcommand{\AveTotStatWp}{   0.02}
\newcommand{\AveTotSystWp}{   0.32}
\newcommand{\AveTotLumiWp}{   0.25}
\newcommand{\AveChisq}{3.0}

\newcommand{\AveTotCentralW}{20.64}
\newcommand{\AveTotStatW}{   0.02}
\newcommand{\AveTotSystW}{   0.55}
\newcommand{\AveTotLumiW}{   0.43}
\newcommand{\AveFidCentralW}{8.03}
\newcommand{\AveFidStatW}{   0.01}
\newcommand{\AveFidSystW}{   0.16}
\newcommand{\AveFidLumiW}{   0.17}

\newcommand{\WZcombVal}{  10.31 } 
 \newcommand{\WZcombSt}{  0.04 } 
 \newcommand{\WZcombSy}{  0.20 } 
\newcommand{\WpWmcombVal}{  1.295 } 
 \newcommand{\WpWmcombSt}{  0.003 } 
 \newcommand{\WpWmcombSy}{  0.010 }

\newcommand{\lumWZ}{\ensuremath{81\;\mathrm{pb}^{-1}}}       

\newcommand{\sigWeplusnolabel}{\ensuremath{ 12.23 \pm 0.03 \pm 0.42 \pm 0.27 }} 
\newcommand{\sigWeminusnolabel}{\ensuremath{ 9.00 \pm 0.02 \pm 0.39 \pm 0.20 }} 

\newcommand{\sigWmuplusnolabel}{\ensuremath{11.75 \pm 0.03 \pm 0.33 \pm 0.27}} 
\newcommand{\sigWmuminusnolabel}{\ensuremath{8.75 \pm 0.02 \pm 0.25 \pm 0.20}} 

\newcommand{\sigfidWeplusnolabel}{\ensuremath{ 4.68 \pm 0.01 \pm 0.14 \pm 0.10}}
\newcommand{\sigfidWeminusnolabel}{\ensuremath{ 3.58 \pm 0.01 \pm 0.14 \pm 0.08}} 

\newcommand{\sigfidWmuplusnolabel}{\ensuremath{4.50 \pm 0.01 \pm 0.09 \pm 0.10}} 
\newcommand{\sigfidWmuminusnolabel}{\ensuremath{3.48 \pm 0.01 \pm 0.08 \pm 0.08}} 

\newcommand{\CWeplus}{\ensuremath{ 0.602\pm0.012}}             
\newcommand{\CWeminus}{\ensuremath{0.614\pm0.012}}            

\newcommand{\CWmuplus}{\ensuremath{ 0.653 \pm 0.012}}            
\newcommand{\CWmuminus}{\ensuremath{0.650\pm0.012}}           

\newcommand{\ntotWeplus}{256,858}     
\newcommand{\ntotWeminus}{206,092}    
\newcommand{\ntotWe}{462,950}      
\newcommand{\ntotWmuplus}{266,592}    
\newcommand{\ntotWmuminus}{208,616}   
\newcommand{\ntotWmu}{475,208}        
\newcommand{\ntotW}{938,158}          
\newcommand{\ntotZ}{79,907}           

\newcommand{\ntotsignalWeplus}{$228060 \pm 510 \pm 4920 $}    
\newcommand{\ntotsignalWeminus}{$177890 \pm 450 \pm 6110 $}    

\newcommand{\ntotsignalWmuplus}{ $237720   \pm 520 \pm 2210   $}   
\newcommand{\ntotsignalWmuminus}{$183180   \pm 460 \pm 2520   $}    

\newcommand{\stotWp}{\mbox{$\sigma_{W^+}^{\rm{tot}}$}}
\newcommand{\stotWm}{\mbox{$\sigma_{W^-}^{\rm{tot}}$}}

\newcommand{\stotWpm}{\mbox{$\sigma_{W^{\pm}}^{\rm{tot}}$}}
\newcommand{\stotZ}{\mbox{$\sigma_{Z}^{\rm{tot}}$}}

\newcommand{\stot}{\mbox{$\sigma^{\rm{tot}}$}}

\newcommand{\sfidWp}{\mbox{$\sigma_{W^+}^{\rm{fid}}$}}
\newcommand{\sfidWm}{\mbox{$\sigma_{W^-}^{\rm{fid}}$}}

\newcommand{\sfidWpm}{\mbox{$\sigma_{W^{\pm}}^{\rm{fid}}$}}
\newcommand{\sfidZ}{\mbox{$\sigma_{Z}^{\rm{fid}}$}}

\newcommand{\sfid}{\mbox{$\sigma^{\rm{fid}}$}}
\newcommand*{\Ztomm}{\ensuremath{Z \rightarrow \mu^{+}\mu^{-}}\xspace}

\newcommand*{\Ztoee}{\ensuremath{Z \rightarrow e^{+}e^{-}}\xspace}

\newcommand*{\Ztoll}{\ensuremath{Z \rightarrow \ell^{+}\ell^{-}}\xspace}

\newcommand*{\Wenp}{\ensuremath{W^+ \rightarrow e^+\nu}\xspace}
\newcommand*{\Wmnp}{\ensuremath{W^+ \rightarrow \mu^+\nu}\xspace}
\newcommand*{\Wenm}{\ensuremath{W^- \rightarrow e^-\overline{\nu}}\xspace}
\newcommand*{\Wmnm}{\ensuremath{W^- \rightarrow \mu^- \overline{\nu}}\xspace}
\newcommand*{\Wtn}{\ensuremath{W \rightarrow \tau\nu}\xspace}

\newcommand*{\mT}{\ensuremath{m_{\text{T}}}\xspace}
\newcommand*{\Wnocharge}{\ensuremath{W}\xspace}

\begin{document}

\maketitle


\clearpage

\section{Introduction}

Measurements of electroweak vector-boson production at hadron colliders provide a
benchmark for the understanding of quantum chromodynamic (QCD) and electroweak (EW)
processes. Predictions are available up to next-to-next-to-leading-order (NNLO) accuracy in
QCD and include EW corrections at next-to-leading-order (NLO) accuracy~\cite{Anastasiou:2003ds}. The cross-section predictions depend on the parton distribution functions (PDFs) and are thus sensitive to the underlying dynamics of strongly interacting particles.
Therefore, measurements of \Wboson and \Zboson-boson\footnote{Throughout this paper, $\Zboson/\gamma^*$-boson production is denoted simply by \Zboson-boson production.}  production offer a unique opportunity to test models of
parton dynamics at the Large Hadron Collider's (LHC)~\cite{1748-0221-3-08-S08001} new higher centre-of-mass energy of $\sqrt{s}=13$~TeV.

This paper describes measurements of the inclusive production cross sections times leptonic branching ratios for
the $W^{\pm}\to e^{\pm}\nu$, $W^{\pm}\to\mu^{\pm}\nu$, \Ztoee, and \Ztomm\ processes.
Measurements of the cross-section ratios of \Wplus to \Wminus production and of \Wboson to \Zboson production are also presented.
All measurements are performed with proton--proton ($pp$) collision data 
corresponding to an integrated luminosity of \lumWZ, 
collected at $\sqrt{s}=13$~TeV with the ATLAS detector~\cite{PERF-2007-01}.
The data were collected during the period of June 13 to July 16, 2015, at which point the LHC circulated
\SI{6.5}{\TeV} beams with \SI{50}{ns} bunch spacing.  The peak
delivered instantaneous luminosity was $L=\SI{1.7e33}{\per \cm \per \s}$
and the mean number of $pp$ interactions per bunch crossing (hard scattering and pile-up events) was $\langle \mu \rangle = 19$.


\section{Methodology of cross-section measurement and predictions}
\label{sec:method}

The total production cross section for the \Wboson boson times the branching ratio for decays into a single-lepton flavour $\ell^{\pm}=e^{\pm},\mu^{\pm}$ (\stotWpm,\stotWp, and \stotWm) can be expressed as a ratio of the numbers of background-subtracted data events $N$ to the product of the integrated luminosity of the data $\mathcal{L}$, an acceptance factor $A$, and a correction factor $C$:
\begin{equation}
\stot = \frac{ N}{\mathcal{L} \cdot A \cdot C}.
\label{EQ::WZxsec}
\end{equation}
The cross sections are defined similarly for the \Zboson boson  in the dilepton invariant mass range $66<m_{\ell\ell}<116$~GeV (\stotZ). 
The acceptance factor $A$ is  expressed as the fraction of decays satisfying the fiducial acceptance (geometric and kinematic requirements) at the Monte Carlo generator level.
The correction factor $C$ is the ratio of the total number of generated events which pass the final selection requirements after reconstruction to the total number of generated events within the fiducial acceptance. This factor,  defined before the decay leptons emit photons via final-state radiation (Born-level FSR),   includes the efficiencies for triggering on, reconstructing, and identifying the \Wboson\ and \Zboson-boson decay products within the acceptance, and also accounts for the slight difference between the fiducial and reconstructed phase spaces. The production cross sections defined without the acceptance factors ($\sigma^{\mathrm{tot}}\cdot A$) are referred to as the fiducial cross sections (\sfidWpm, \sfidWp, \sfidWm, and \sfidZ). For the \Wboson-boson measurement, the fiducial phase space is defined by the lepton transverse momentum $\pT^{\ell}>25$~GeV, the lepton pseudorapidity\footnote{ATLAS uses a right-handed coordinate system with its origin at the nominal 
interaction point (IP) in the centre of the detector and the
$z$-axis along the beam pipe. The $x$-axis points from the IP to the
centre of the LHC ring, and the $y$-axis points upward. Cylindrical
coordinates ($r$, $\phi$) are used in the transverse plane, $\phi$
being the azimuthal angle around the beam pipe. The pseudorapidity
is defined in terms of the polar angle $\theta$ as $\eta = -\ln \tan(\theta/2)$.}
 $|\eta_{\ell}|<$~2.5, the neutrino transverse momentum $\pT^{\nu}>25$~GeV, and the \Wboson-boson transverse mass\footnote{$\mT = \sqrt{ 2 \pT^{\ell}\ \pT^{\nu} \left[1 - \cos{(\phi_{\ell}-\phi_{\nu})  } \right] }$
with azimuthal angle of the charged lepton $\phi_{\ell}$ and azimuthal angle of the neutrino $\phi_{\nu}$.} $\mT>50$~GeV.  
Similarly, \Zboson-boson production is measured in the fiducial phase space defined by $\pT^{\ell}>25$~GeV, $|\eta_{\ell}|<$~2.5, and  $66<m_{\ell\ell}<116$~GeV.


Theoretical predictions of the fiducial and total cross sections are computed using {\sc DYNNLO}~1.5~\cite{Catani:2007vq,Catani:2009sm} for the central value and {\sc Fewz}~3.1~\cite{Anastasiou:2003ds,FEWZ2,FEWZ3,FEWZ4} for all variations reflecting systematic uncertainties, thereby providing full NNLO QCD calculations. The NLO EW corrections are calculated with {\sc Fewz}~3.1 for \Zboson bosons and with the Monte Carlo program {\sc Sanc}~\cite{Bardin:2012jk,Arbuzov:2012dx} for \Wboson bosons.  The calculation is done in the $G_\mu$ EW scheme~\cite{Hollik:1988ii}. The cross sections are calculated for vector-boson decays into leptons at Born level, to match the definition of the $C$ factor used in Eq.~\ref{EQ::WZxsec} for the determination of the measured cross sections in the data. Thus, from complete NLO EW corrections the following components are included: virtual QED and weak corrections, initial-state radiation (ISR) and interference between ISR and FSR~\cite{Dittmaier:2009cr}. 
For the \Zboson-boson production, all the predictions include the $66<m_{\ell\ell}<\SI{116}{\GeV}$ requirement.
The NNLO PDFs  {\sc CT14}nnlo~\cite{Dulat:2015mca}, {\sc NNPDF3.0}~\cite{Ball:2014uwa}, {\sc MMHT14}nnlo68CL~\cite{Harland-Lang:2014zoa}, {\sc ABM12}~\cite{Alekhin:2013nda}, {\sc HERAPDF2.0}nnlo~\cite{Abramowicz:2015mha}, 
and ATLAS-epWZ12nnlo \cite{Aad:2012sb} are used in the comparisons to data, although {\sc CT14}nnlo is used as the baseline for the predictions.

The systematic uncertainties in the predictions are dominated by the imperfect knowledge of the proton parton distribution functions. These uncertainties are obtained from the sum in quadrature of the differences between the central PDF values and the eigenvectors of the respective PDF sets.  Where appropriate, asymmetric uncertainties are determined using separate sums of negative and positive variations. The {\sc CT14}nnlo uncertainties (rescaled from 90\%  to 68\% confidence level (CL)) are used in the comparison to the measured cross sections in Table~\ref{tab:pred_meas_xsections} of Section~\ref{sec:results}. 
The QCD scale uncertainties are defined by the symmetrised envelope of variations in which the renormalisation ($\mu_{\rm R}$) and factorisation ($\mu_{\rm F}$) scales are changed by factors of two with an additional constraint of $0.5\leq \mu_{\rm R} / \mu_{\rm F} \leq2$. The dynamic scale $m_{\ell\ell}$ and fixed scale $m_{W}$ are used as the central values for the \Zboson boson and \Wboson boson predictions, respectively. A significant component of these scale uncertainties originates from the statistical precision of the integration method used to evaluate the variations.
The other systematic uncertainties under consideration (labelled as ``other'' in Table~\ref{tab:pred_meas_xsections}) are as follows. The uncertainties due to the strong coupling constant are estimated following the prescription given with the {\sc CT14}nnlo PDF, varying $\alpha_{\rm S}$ by $\pm 0.001$ to correspond to $68\%$ CL. The beam energy is assumed to be known to  1\%~(from Ref.~\cite{TOPQ-2012-05}, with an additional uncertainty to take into account the extrapolation of this uncertainty to 13~TeV). 
The limitations of the NNLO calculations are estimated by comparing the predictions calculated with {\sc DYNNLO}~1.5 and with {\sc Fewz}~3.1. For the total cross-section predictions, these differences are found to be $<0.2$\% per process and hence are negligible. For the fiducial cross-section predictions, these differences are larger due to a feature of the calculations involving leptons with symmetric \pT requirements, resulting in consistently larger values from {\sc Fewz}. The differences are calculated using the {\sc CT14}nnlo PDF as a central value in both cases, and are up to 1.3\% for the \Wboson-boson cross sections and 0.6\% for the \Zboson-boson cross section.  These differences are however not included in the prediction uncertainties given in Table~\ref{tab:pred_meas_xsections} of Section~\ref{sec:results}.

Predictions for the fiducial cross-section ratios $\sfidWp/\sfidWm$ and $\sfidWpm/\sfidZ$ are also calculated, with their corresponding PDF uncertainties considered as fully correlated, eigenvector by eigenvector, in the ratios. The QCD scale variations are not considered for the ratios since the higher-order corrections are expected to affect both the \Wboson and \Zboson bosons in a similar manner but the exact correlation is difficult to evaluate. The differences between {\sc Fewz} and {\sc DYNNLO} for \Wplus/\Wminus and \Wboson/\Zboson  are 0.4\% and 0.6\%, respectively, and are not included in the prediction uncertainties of Table~\ref{tab:pred_meas_xsections}. The remaining theoretical uncertainties evaluated in the fiducial cross sections mentioned above largely cancel in the ratio and are also neglected.

The acceptance factors $A$ are also calculated with {\sc DYNNLO}~1.5 for the central value and {\sc Fewz}~3.1 for variations accounting for systematic uncertainties.  Their uncertainties are derived from the envelope of the PDF variations of  {\sc CT14}nnlo, {\sc NNPDF3.0}, {\sc MMHT14}nnlo68CL, and {\sc ABM12}. Calculations of the acceptance factors obtained from either the signal Monte Carlo simulation used in this analysis (\POWHEG+~\PYTHIA~8~\cite{Nason:2004rx,Frixione:2007vw,Alioli:2010xd,Sjostrand:2007gs}, fully described in Section~\ref{sec:mc}) or from {\sc Fewz} fall within this envelope.
In addition, uncertainties due to parton showers and the hadronisation description are taken from a previous publication~\cite{Aad:2011dm}, after checking their validity for the 13~TeV result, and  were derived as the difference in the acceptances calculated with \POWHEG-Box~v1 but using different models for parton shower and hadronisation
descriptions, namely the \HERWIG~\cite{Corcella:2000bw} or \PYTHIA~\cite{Sjostrand:2006za} programs.


\section{Simulation samples}
\label{sec:mc}

Monte Carlo simulations are used to evaluate the selection efficiency for signal events and the contribution of several background processes to the analysed dataset.
All of the samples are processed with the {\textsc{Geant}4}-based
simulation~\cite{Agostinelli:2002hh} of the ATLAS detector~\cite{SOFT-2010-01}.
Nearly all of the processes of interest, specifically events containing \Wboson or \Zboson bosons~\cite{Alioli:2008gx}, are generated with the \POWHEG-Box~v2 Monte Carlo
program interfaced to the \PYTHIA~8.186 parton shower model.
The CT10 PDF set~\cite{Lai:2010vv} is used in the matrix element and the AZNLO~\cite{STDM-2012-23} set of generator-parameter values (tune) is used, with PDF set CTEQ6L1~\cite{Pumplin:2002vw}, for the modelling of non-perturbative effects.
The EvtGen v.1.2.0 program~\cite{EvtGen} is used for properties of the bottom and charm hadron decays, and
{\sc Photos}++ version 3.52~\cite{Golonka:2005pn,Davidson:2010ew} is used for QED emissions from electroweak vertices and charged leptons.
Samples of top-quark pair ($t\overline{t}$) and single-top-quark production are generated with the \POWHEG-Box v2 generator, which uses the four-flavour scheme for the NLO matrix element calculations together with the fixed four-flavour PDF set CT10f4. For all top-quark processes, top-quark spin correlations are preserved.
The parton shower, fragmentation, and underlying event are simulated using \PYTHIA~6.428 with the CTEQ6L1 PDF sets and the corresponding Perugia 2012 tune (P2012)~\cite{Skands:2010ak}. The top-quark mass is set to 172.5 GeV. The EvtGen v1.2.0 program is used for properties of the bottom and charm hadron decays.
Diboson processes are simulated using the \SHERPA~v2.1.1 generator~\cite{Gleisberg:2008ta}.
Multiple overlaid $pp$ collisions are simulated with the soft QCD processes of \PYTHIA~v.8.186 using the A2 tune~\cite{ATLAS-PHYS-PUB-2012-003} and the {\sc MSTW2008LO} PDF~\cite{Martin:2009iq}.
The Monte Carlo events are reweighted
so that the $\langle \mu \rangle$ distribution matches the observed pile-up distribution in the data.
For the comparison to data in the distributions, the single-boson Monte Carlo simulations are normalised to the cross sections measured by this analysis. In the evaluation of the single-boson EW backgrounds for the cross-section calculations, simulations are instead normalised to the results of higher-order QCD calculations, with uncertainties of 5\%. The remaining simulations are also normalised to the predictions of higher-order QCD calculations, with uncertainties of 6\% for the diboson and top-quark processes.


\section{Event selection}

Electron and muon candidate events are selected using triggers which require at least one electron or muon with
transverse momentum thresholds of $\pT=\SI{24}{\GeV}$ or \SI{20}{\GeV}, respectively, with loose
isolation requirements. To recover possible efficiency losses at high momenta, additional electron and muon triggers which do not make any isolation requirements are included with thresholds of $\pT=\SI{60}{\GeV}$ and \SI{50}{\GeV}, respectively. 

Electron candidates are required to have $\pT>\SI{25}{\GeV}$ and to pass the ``medium'' likelihood-based identification requirements~\cite{elecperf,ATL-PHYS-PUB-2015-041} optimised for the 2015 operating conditions, within the fiducial region of $|\eta|<2.47$, excluding candidates in the transition region between the
barrel and endcap electromagnetic calorimeters,
$1.37<|\eta|<1.52$. Muons are reconstructed for $|\eta|<2.4$ with $\pT>\SI{25}{\GeV}$ and must pass the ``medium'' identification requirements~\cite{Aad:2016jkr} also optimised for the 2015 operating conditions. At least one of the lepton candidates is required to match the lepton that triggered the event.
The electrons and muons must also satisfy \pT-dependent cone-based isolation requirements,
using both tracking detector and calorimeter information (described in Refs.~\cite{Aad:2015yja,Aad:2014lwa}, respectively). The isolation requirements are
tuned so that the lepton isolation efficiency is at least 90\% for all $\pT>\SI{25}{\GeV}$, increasing to 99\% at 60~GeV.

Jets are reconstructed from energy deposits in the calorimeter using the anti-$k_t$
algorithm~\cite{Cacciari:2008} with radius parameter $R=0.4$.
All jets~\cite{ATL-PHYS-PUB-2015-015}, with energies calibrated at the electromagnetic scale, must have $\pT>\SI{20}{\GeV}$ and $|\eta|<4.5$. 
The missing transverse momentum (with magnitude \MET), which in the \Wboson-boson analysis acts as a proxy for the transverse momentum of the neutrino,  
is defined as the negative of the global vector sum of all identified physics objects (electrons, muons, jets)
as well as specific ``soft terms'' accounting for unclassified soft tracks and calorimeter energy clusters.

The event selection for the \Wboson-boson signature requires exactly one identified
electron or muon. The event is required to have $\MET>\SI{25}{\GeV}$, and the transverse mass of the \Wboson boson calculated using the  missing transverse momentum vector
is required to satisfy $\mT>\SI{50}{\GeV}$.
In order for the \Wboson-boson selection to be consistent with the missing transverse momentum reconstruction methodology, an overlap removal algorithm is applied to the selection for events with jets and leptons found at a distance of $\Delta R = \sqrt{(\Delta \eta)^2 + (\Delta \phi)^2} < 0.4$ of each other, removing either one or the other object. 
After the full \Wln\ selection, a total of \ntotWe~\Wboson-boson candidates (\ntotWeplus~\Wplus and \ntotWeminus~\Wminus) 
pass all requirements in the electron channel, and \ntotWmu~\Wboson-boson candidates (\ntotWmuplus~\Wplus and \ntotWmuminus~\Wminus)
pass the requirements in the muon channel.

Events containing a \Zboson-boson candidate are selected by requiring exactly two selected leptons
of the same flavour but of opposite charge with invariant mass of  $66 < m_{\ell\ell} < \SI{116}{GeV}$. No overlap removal is applied in the \Zboson-boson analysis, as missing transverse momentum is not required in the selection.
A total of \ntotZe~candidates pass all requirements in the electron channel and \ntotZmu~candidates in the muon channel.

\section{Evaluation of backgrounds}

Contributions from the electroweak (single-boson and diboson) and top-quark (single-top and top-quark pair) components of the background are estimated from the Monte Carlo samples described earlier. The \Wtn and \Ztautau processes with the subsequent leptonic decay of the $\tau$  are treated as background. The dominant contributions, given as percentages of the total number of simulated events passing the signal selection in each analysis, are as follows: the \Wtn and top-quark production contribute approximately 2\% and 1\%, respectively, in the \Wboson-boson analyses, the \Ztoee\ and \Ztomm\ processes contribute 1\% and 5\% in \Wen\ and \Wmn, respectively, while the total background in \Ztoll\ is approximately 0.5\%, dominated by $t\overline{t}$ production (the sum of all electroweak backgrounds is 0.2\%).  

Events involving semileptonic decays of heavy quarks, hadrons misidentified as leptons, and, in the case of the electron channel, electrons from photon conversions (all referred to collectively as ``multijet events'') are a sizeable source of background in the \Wboson-boson analysis.
The multijet background in the \Zboson-boson analysis is estimated from simulation to be < 0.1\% and is therefore neglected.

The multijet contribution to the electron and muon channels of the \Wboson-boson analysis is estimated with a data-driven approach, performing maximum-likelihood fits on the data with template distributions to exploit the discriminating power between signal and background in certain kinematic distributions. The discriminant variables used in the multijet evaluation are \mT, \MET, $\pT^{\ell}$, and $\Delta \phi$ between the lepton and transverse missing momentum.
Two fit regions are used to extract the multijet normalisation. The first fit region is defined as the full event selection but removing the \mT requirement, and the second one is defined as the full event selection but removing the \MET requirement. 
Several multijet-enriched data samples (multijet templates) are built from events passing all selection requirements in each fit region except lepton isolation.
Mutually exclusive requirements (``intervals'') in either tracking- or calorimeter-based isolation variables are chosen to create statistically independent multijet templates. These samples are designed to be progressively closer to the signal-candidate selection by fixing one of the isolation criteria to that of the signal region and varying the other one; four such samples are built for each isolation type in the electron channel and four (for tracking-based isolation) or six (for calorimeter-based isolation) in the muon channel.
Templates are similarly constructed from simulation for \Wboson signal and electroweak and top-quark backgrounds, to account for potential contaminations in the multijet template.
For each isolation interval, the normalisation of the multijet template is extracted with a maximum-likelihood fit to the data in the two fit regions and separately for each one of the discriminant variables and charged lepton samples.
In each fit region, the normalisation of the signal template derived from simulation is left free to float while the remaining background templates are normalised to their expected values, based on the measured integrated luminosity and the predicted cross sections (but are permitted to vary within 5\% of their expected values, as described in Section~\ref{sec:mc}). It was verified that the value of the signal normalisation extracted from this fit has no significant impact on the multijet estimate.

\begin{figure}[t]
  \centering
  \includegraphics[width=0.48\textwidth]{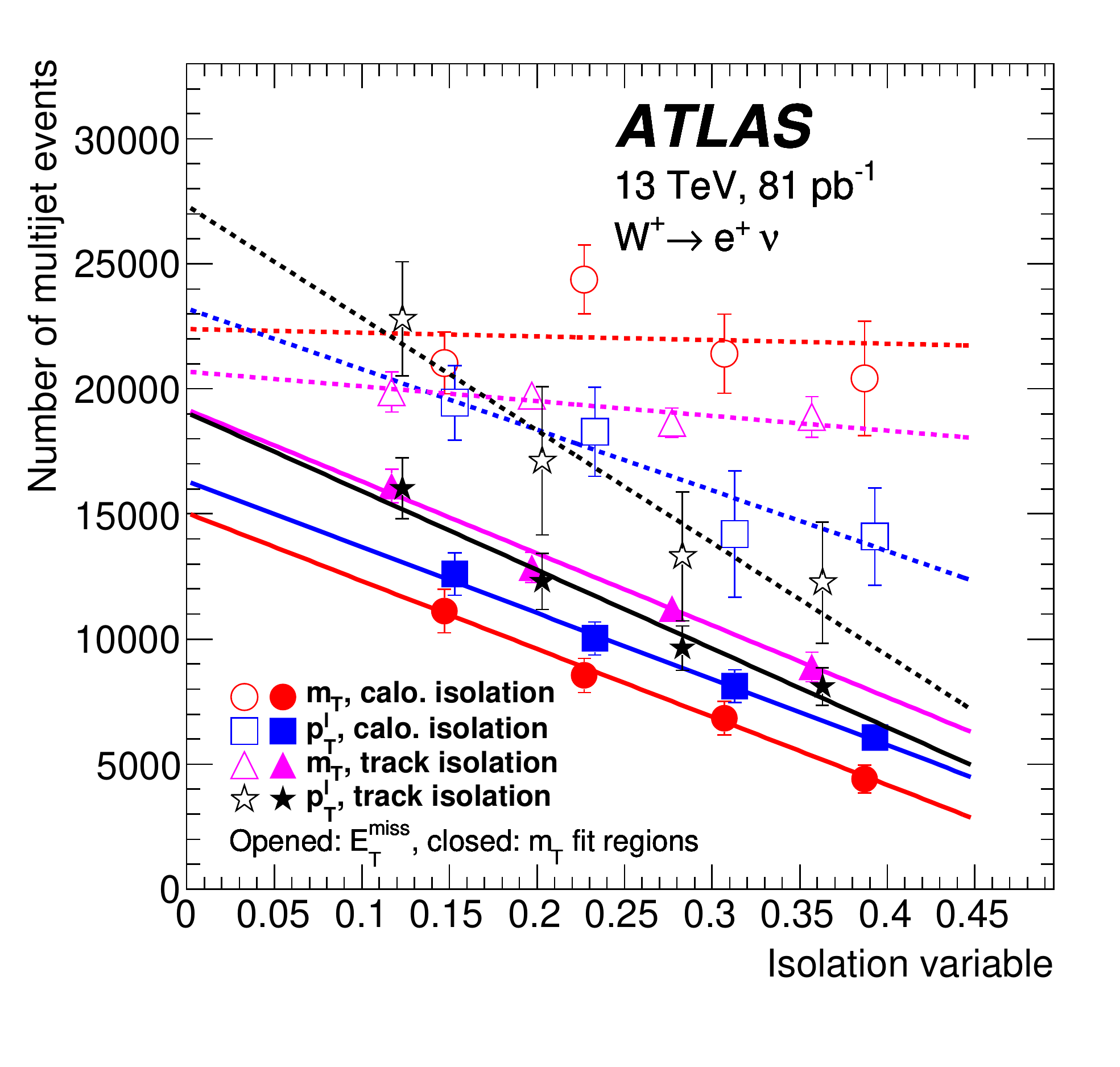}
  \includegraphics[width=0.48\textwidth]{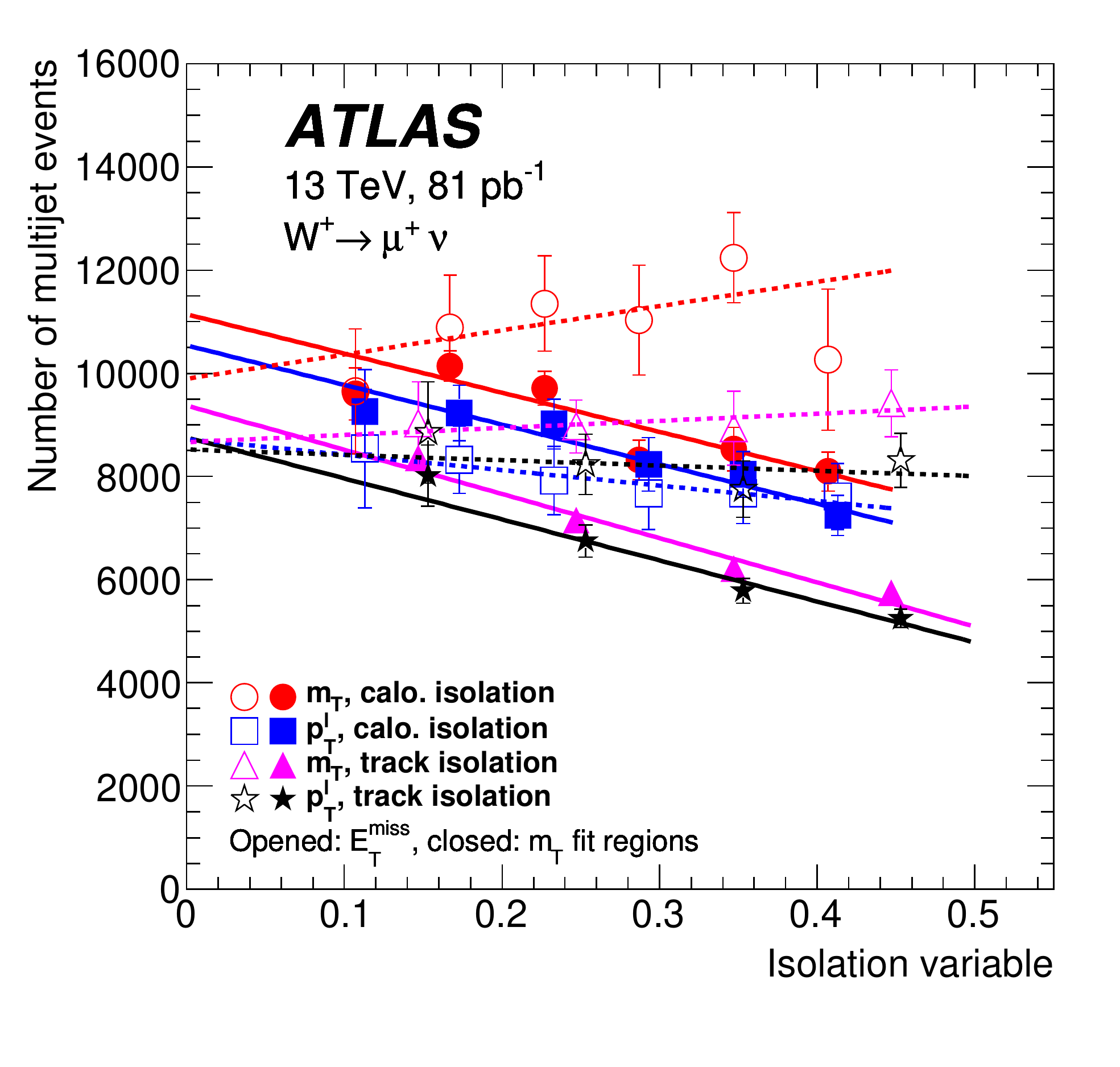}
\caption{The number of multijet events versus the isolation variable for the  \Wen (left) and \Wmn (right) analysis is shown. The plots illustrate the multijet-evaluation methodology for the \Wplus\ analysis. The results obtained for two of the four discriminant variables used to evaluate the multijet yields are shown for both types of isolation: \mT (circles) and $\pT^{\ell}$ (squares) with calorimeter-based isolation and \mT (triangles) and $\pT^{\ell}$ (stars) with track-based isolation. Open markers represent the yields obtained with the \met fit region while closed markers are those with the \mT fit region. The points represent the extracted multijet fraction from the fit of the variables, in the isolation intervals represented on the $x$-axis for the template selection. The lines represent the linear extrapolation of the points to the signal region. The definition of the signal region is \pT and isolation-flavour dependent but corresponds approximately to the region of isolation below 0.1 in these plots. The error bar in each bin represents the uncertainty from the fit of the variable rescaled by the square root of the reduced $\chi^2$ of the fit.}
  \label{fig:isoscan}
\end{figure}

The multijet background event yield in each region is then estimated from this normalisation together with the signal-region requirement of either $\mT>\SI{50}{\GeV}$, or $\MET>\SI{25}{\GeV}$. 
For each discriminant variable, and separately for calorimeter- and track-based isolation and for each fit region, the
estimates obtained in the isolation intervals are used to build a linear extrapolation to the isolation selection used
in the signal region. The extrapolation is performed assuming that the individual estimates are
uncorrelated.
Figure~\ref{fig:isoscan} illustrates this multijet-evaluation methodology for the \Wplus\ analysis using two of the variables, \mT and $\pT^{\ell}$, and for both fit regions.

Separately for the calorimeter- and track-based isolation variables, and for each fit region, an estimate of the background yield is obtained from a weighted average of the extrapolated values obtained with the individual discriminant variables and their uncertainties, after verifying their compatibility with a $\chi^2$ criterion.
The average of the four multijet background estimated fractions found from the track and calorimeter isolation requirements in each fit region is then taken as the nominal multijet background yield in each channel.
The uncertainties derived from the linear extrapolations are propagated as systematic uncertainties in the method. A systematic uncertainty for the choice of isolation variable is obtained from half the difference between the averages of the calorimeter-based isolation estimated fractions in the two fit regions, and the track-based averages. Similarly, a systematic uncertainty due to the use of different fit regions is evaluated as half the difference of the averages obtained from the different types of isolation in the two separate fit regions. In addition, the impact of variations of the jet-energy scale on the signal template is added in quadrature to the multijet systematic uncertainty.

The estimated multijet background fractions of the total number of observed candidate events are 8\% and 10\% in the electron \Wplus\ and \Wminus\ channels  and 3.5\% and 4\% in the muon \Wplus\ and \Wminus channels. The corresponding relative uncertainties range from approximately 20\% to 30\%  for the muon and electron channels, and are similar for both the positively and negatively charged samples.


\section{Evaluation of systematic uncertainties} 

The experimental systematic uncertainties in the measurements of the cross sections enter via the evaluation of the correction factor and the luminosity in the denominator of Eq.~(\ref{EQ::WZxsec}), and through the estimation of the background subtracted from the candidate events in its numerator.

The sources of systematic uncertainties in the correction factors $C$, summarised in Table~\ref{tab:systotWZ}, are as follows. 
{\it Trigger:} The lepton trigger efficiency is estimated in simulation, with a dedicated data-driven analysis
  performed to obtain the simulation-to-data trigger correction factors and the corresponding uncertainties. 
{\it Reconstruction, Identification, and Isolation:} The lepton selection efficiencies as determined from simulation are corrected with simulation-to-data correction factors and their associated uncertainties~\cite{ATL-PHYS-PUB-2015-041,Aad:2016jkr}. 
{\it Energy, Momentum Scale/Resolution:} Uncertainties in the lepton calibrations are applied as they can cause a change of acceptance because of migration of events across the \pT\ threshold and $m_{\ell\ell}$ boundaries. 
{\it Charge Identification:} Electron charge misidentication may occur when electrons radiate early in the detector  and the resulting photons
subsequently convert and are reconstructed as high \pT\ tracks. A particle with reconstructed charge opposite to the parent electron may then accidentally be associated with the calorimeter cluster. The effect of electrons having their charge reconstructed wrongly is studied using a control sample of $Z\rightarrow ee$ events in which both electrons are reconstructed with the same charge and is found to be well described by the Monte Carlo simulation, within the statistical uncertainty of the control sample. An uncertainty is assessed to cover any small residual differences between data and simulation. The probability of charge misidentification is negligible in the muon channel.
{\it Jet-Energy Scale/Resolution (JES and JER):} The corresponding uncertainties, described in Ref.~\cite{ATL-PHYS-PUB-2015-015}, are propagated to the calculation of the missing transverse momentum.
{\it \MET:} Uncertainties in the soft component of the \MET~scale and resolution evaluated as described in Ref.~\cite{ATL-PHYS-PUB-2015-04} are included. 
{\it Pile-up:} Incorrect modelling of pile-up effects can lead to acceptance changes and is accounted for with dedicated studies.  
{\it PDF:} The impact of PDF eigenvector variations is propagated to the correction factor.

In the \Zboson-boson channel, the systematic uncertainties from the background evaluation contribute negligibly to the experimental cross-section uncertainty. This is not the case for the \Wboson-boson channel; the multijet background evaluation results in uncertainties of up to 3.4\% on the cross-section measurements in the electron channel and up to 1.4\% in the muon channel.

The measurement of the integrated luminosity has a 2.1\% uncertainty, which is derived, following a methodology similar to that detailed in Ref.~\cite{DAPR-2011-01}, from a calibration of the luminosity scale using $x$--$y$ beam-separation scans  performed in August 2015. Apart from the determination of the luminosity, the dominant experimental systematic uncertainties in the cross-section evaluations are the jet-energy scale/resolution and the multijet background for the \Wboson-boson measurements while they are lepton reconstruction and identification efficiencies for the \Zboson-boson measurements.

\renewcommand{\tabcolsep}{1.5pt}

\begin{table}[t]
\begin{center}
\footnotesize
\begin{tabular}{l||ccc|ccc}
\hline
\hline
$\delta C/C$ [\%]  &\Ztoee\ & \Wenp\ & \Wenm\ & \Ztomm\ & \Wmnp\ & \Wmnm\ \\
\hline
Lepton trigger                         &            0.1 &    0.3 &    0.3 &    0.2 &    0.6 &    0.6 \\ 
Lepton reconstruction, identification  &            0.9 &    0.5 &    0.6 &    0.9 &    0.4 &    0.4  \\ 
Lepton isolation                       &            0.3 &    0.1 &    0.1 &    0.5 &    0.3 &    0.3  \\ 
Lepton  scale and resolution           &            0.2 &    0.4 &    0.4 &    0.1 &    0.1 &    0.1  \\ 
Charge identification                  &            0.1 &    0.1 &    0.1 &  --  &  --  &  --         \\ 
JES and JER                            &            --  &    1.7 &    1.7 &  --  &    1.6 &    1.7    \\ 
\MET                                   &            --  &    0.1 &    0.1 &  --  &    0.1 &    0.1    \\ 
Pile-up modelling                      &         $<0.1$ &    0.4 &    0.3 &    $<0.1$ &    0.2 &    0.2  \\ 
PDF                                    &         0.1 &    0.1 &    0.1 &    $<0.1$ &    0.1 &    0.1  \\ 
\hline 
Total                                  &            1.0 &    1.9 &    1.9 &    1.1 &    1.8 &    1.8  \\ 

\hline 
\hline
\end{tabular}
\end{center}
\caption{Relative systematic uncertainties (\%) in the correction factors $C$ in the different channels. 
\label{tab:systotWZ}}
\end{table}
\renewcommand{\tabcolsep}{6pt}

\section{Results}
\label{sec:results}

\begin{figure}[t]
  \centering
  \includegraphics[width=0.48\textwidth]{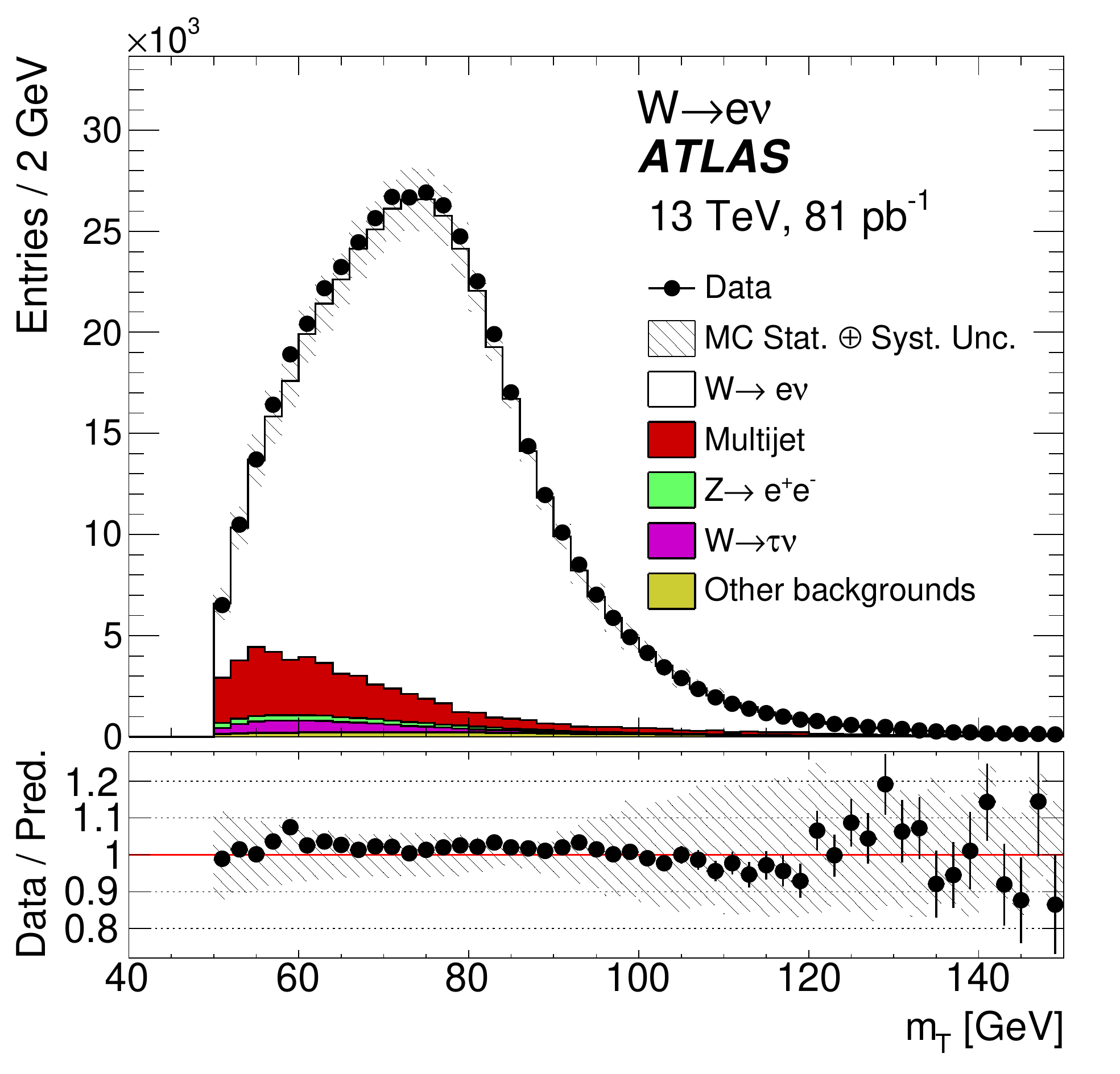}
  \includegraphics[width=0.48\textwidth]{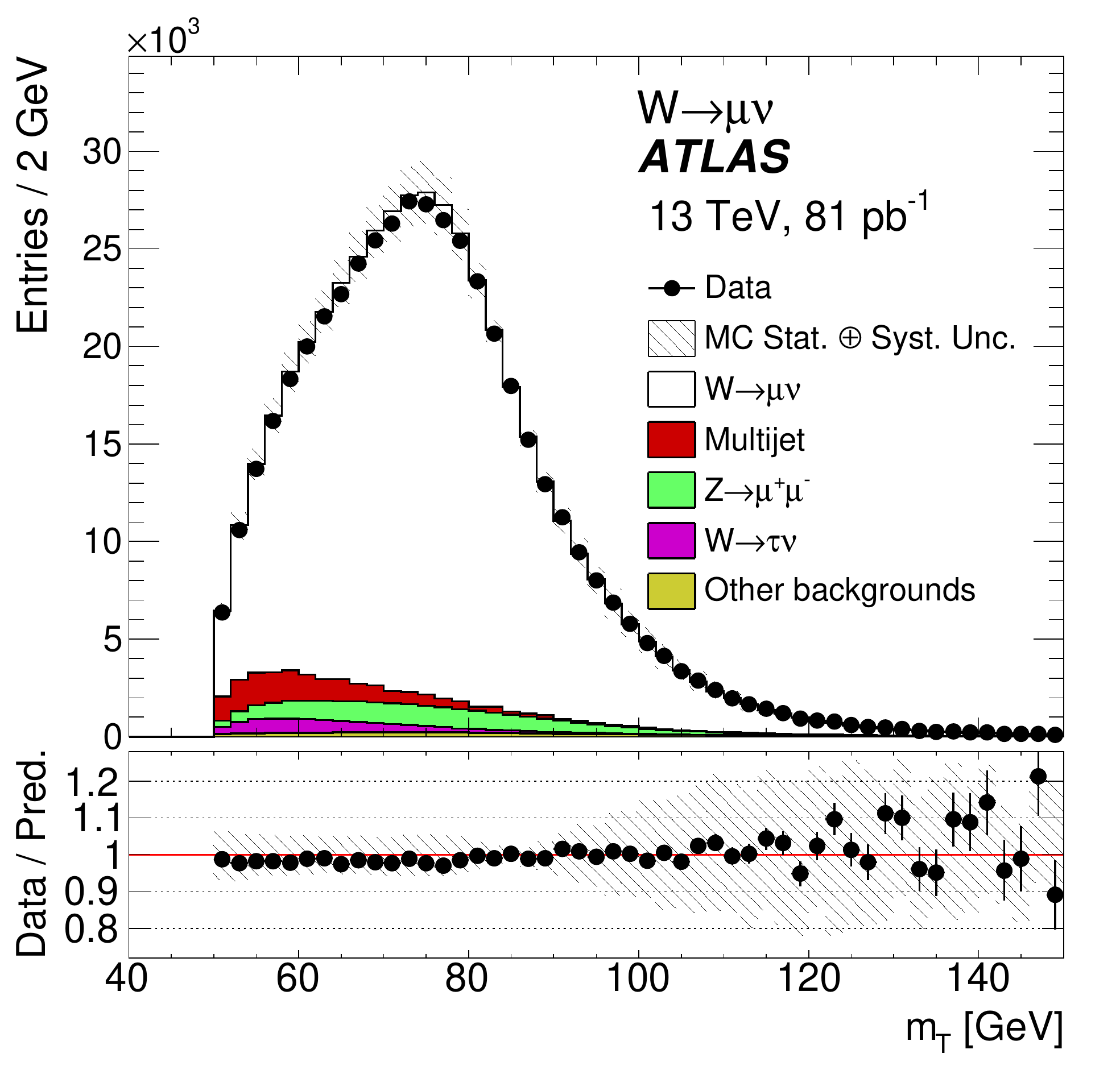}
  \includegraphics[width=0.48\textwidth]{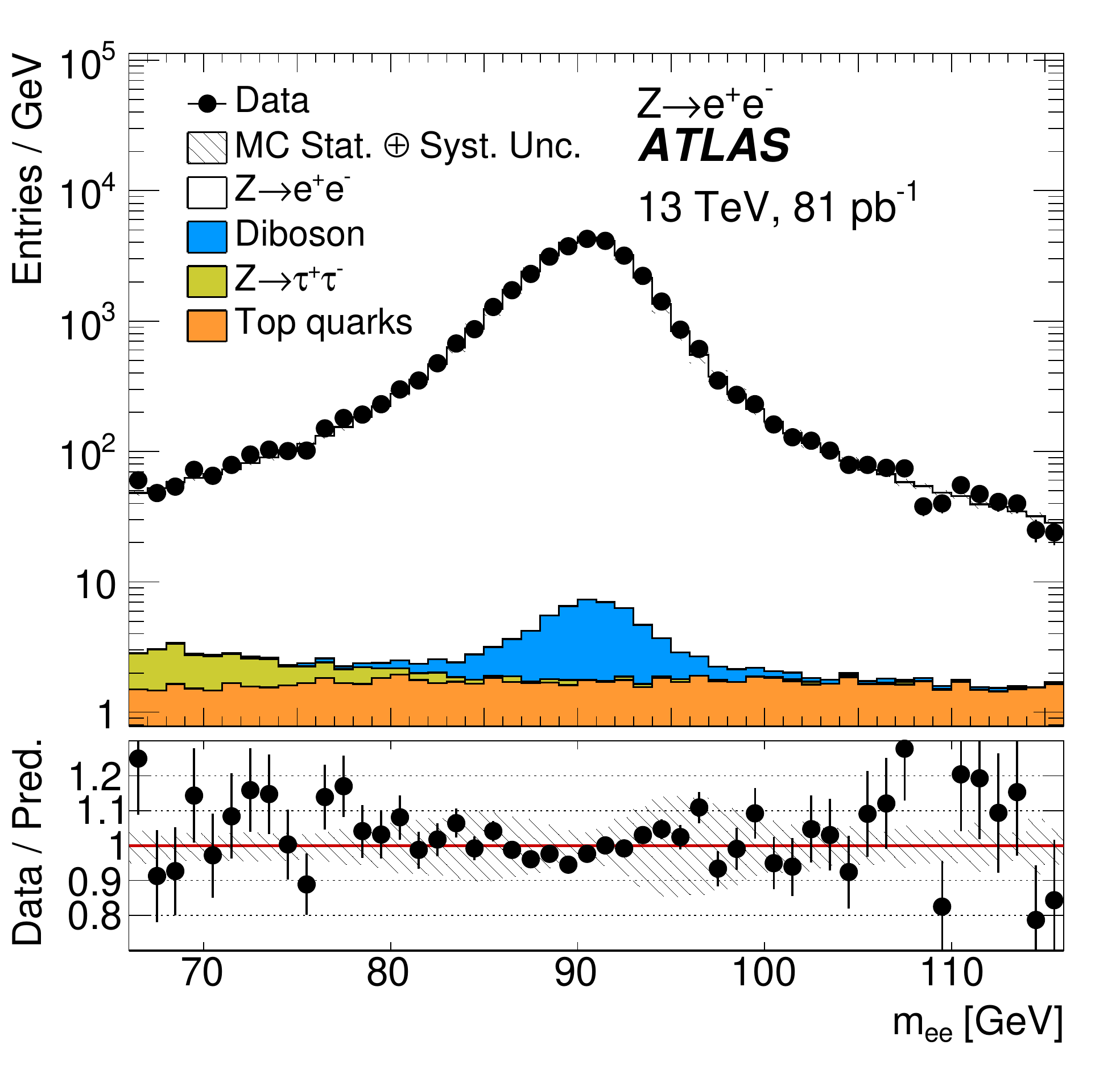}
  \includegraphics[width=0.48\textwidth]{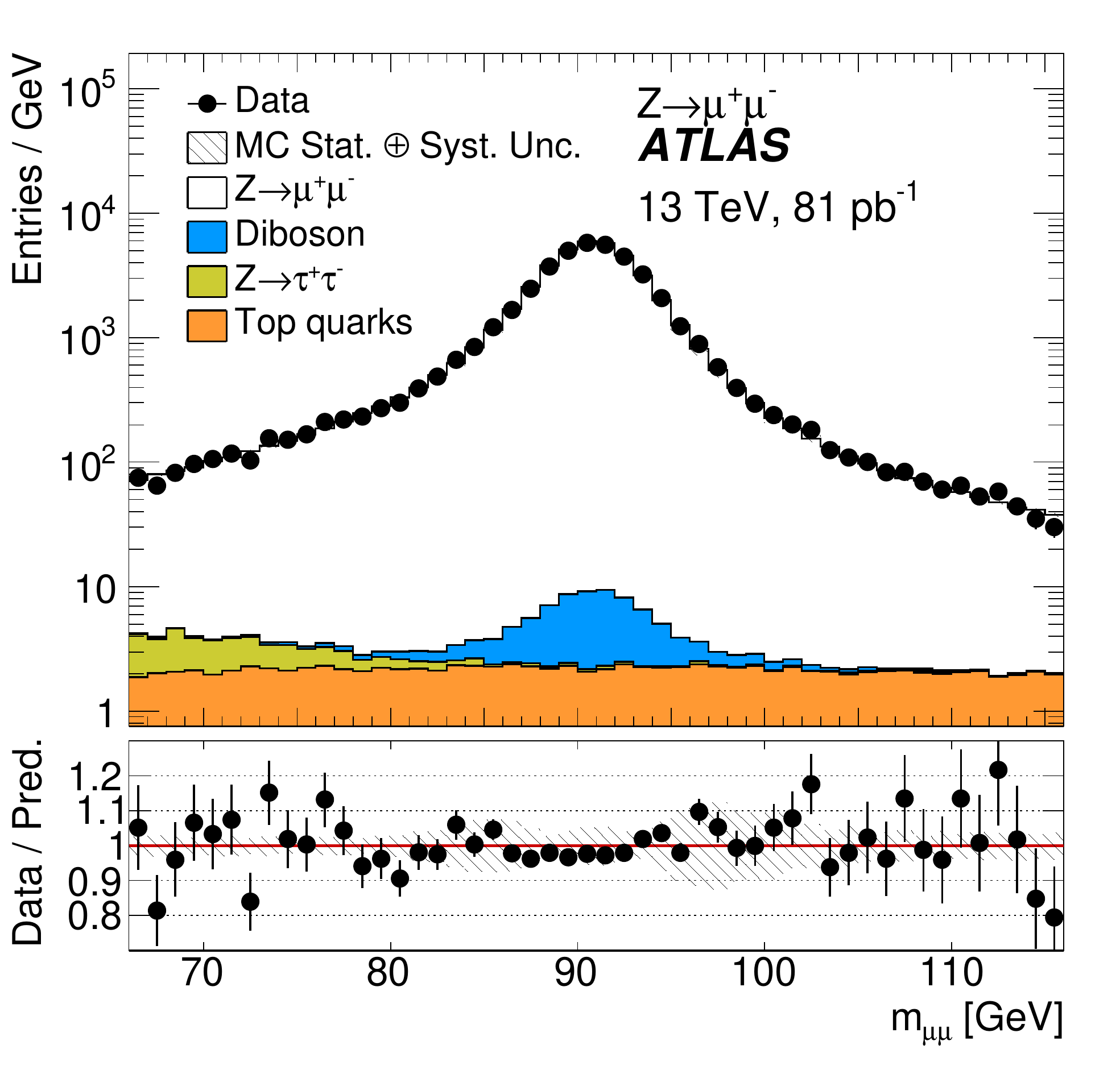}
  \caption{Transverse mass distributions from the \Wen\ and  \Wmn\ selections (top) and dilepton mass distributions from the \Ztoee\ and \Ztomm\ selections (bottom). The predicted signal distributions are normalised to the measured cross sections as presented in this paper.
The shaded bands in the histograms encompass the uncertainties described in Table~\ref{tab:systotWZ}. In addition to these uncertainties in the correction factors, the uncertainties in the evaluation of the multijet background in the \Wboson-boson analysis are included in the shaded bands.}
  \label{fig:kinematics_dist}
\end{figure}

The \mT\ and $m_{\ell\ell}$ distributions after the final selection are shown in Figure~\ref{fig:kinematics_dist}  for the \Wen, \Wmn and \Ztoee, and \Ztomm\ channels, respectively, for the data compared to the predictions, normalised to the measured cross section. All elements necessary to calculate the cross sections for \Wplus, \Wminus and \Zboson-boson production and decay in the electron and muon channels are summarised in Table~\ref{FiducialWZXSections}. The derived fiducial and total cross sections are also presented in this table, along with their statistical, systematic, and luminosity uncertainties.

\begin{table}[t]
\begin{center}
\footnotesize 
\begin{tabular}{|l||r|r|r|}
\cline{1-4}
   & \multicolumn{1}{|c|}{$W^+$} & \multicolumn{1}{|c|}{$W^-$} & \multicolumn{1}{|c|}{$Z$}  \\ \hline \hline
& \multicolumn{3}{|c|}{\bf Electron channel (value $\pm$ stat $\pm$  syst $\pm$ lumi)}  \\ \hline \hline
    Signal events     &  \ntotsignalWeplus $\ \pm\ 200$ & \ntotsignalWeminus $\ \pm\ 180$ &  \ntotsignalZe \\ \hline
    Correction $C$ & \CWeplus  & \CWeminus  & \CZe $ \begin{smallmatrix} \CZeSystUpErr \\ \CZeSystDownErr \end{smallmatrix}$ \\ \hline
    \sfid [nb]  & \sigfidWeplusnolabel & \sigfidWeminusnolabel & \sigfidZenolabel  \\   \hline \hline
    Acceptance $A$ &  
0.383 $\pm$ 0.007 &
0.398 $\pm$ 0.007 &
0.393 $\pm$ 0.007 \\
\hline
    \stot [nb]     & \sigWeplusnolabel & \sigWeminusnolabel & \sigZenolabel \\ \hline \hline
& \multicolumn{3}{|c|}{\bf Muon channel (value $\pm$ stat $\pm$  syst $\pm$ lumi)}   \\\hline \hline
    Signal events     &  \ntotsignalWmuplus $\ \pm\ 410$ & \ntotsignalWmuminus $\ \pm\  360$ & \ntotsignalZmu \\ \hline
    Correction $C$  & \CWmuplus & \CWmuminus  & \CZmu  $\pm {\CZmuSyst}$  \\ \hline
    \sfid [nb] & \sigfidWmuplusnolabel & \sigfidWmuminusnolabel &  \sigfidZmunolabel  \\   \hline \hline
    Acceptance $A$ &
0.383 $\pm$  0.007 &
0.398 $\pm$  0.007 &
0.393 $\pm$  0.007 \\
\hline
    \stot [nb] & \sigWmuplusnolabel & \sigWmuminusnolabel & \sigZmunolabel \\
 \hline
\end{tabular}
\caption{The fiducial and total  cross sections for \Wplus, \Wminus, and \Zboson bosons
in the electron and muon channels.  The observed numbers of signal events after background subtraction are shown for each channel, along with the correction factors~$C$ and  the geometrical acceptance correction factors~$A$ (both given with the total uncertainty only). The luminosity uncertainties in the measured number of signal events correspond to those from the electroweak and top-quark backgrounds estimated from simulation. 
\label{FiducialWZXSections}} 
\end{center}
\end{table}

\begin{figure}[t]
\centering
\includegraphics[width=0.60\textwidth]{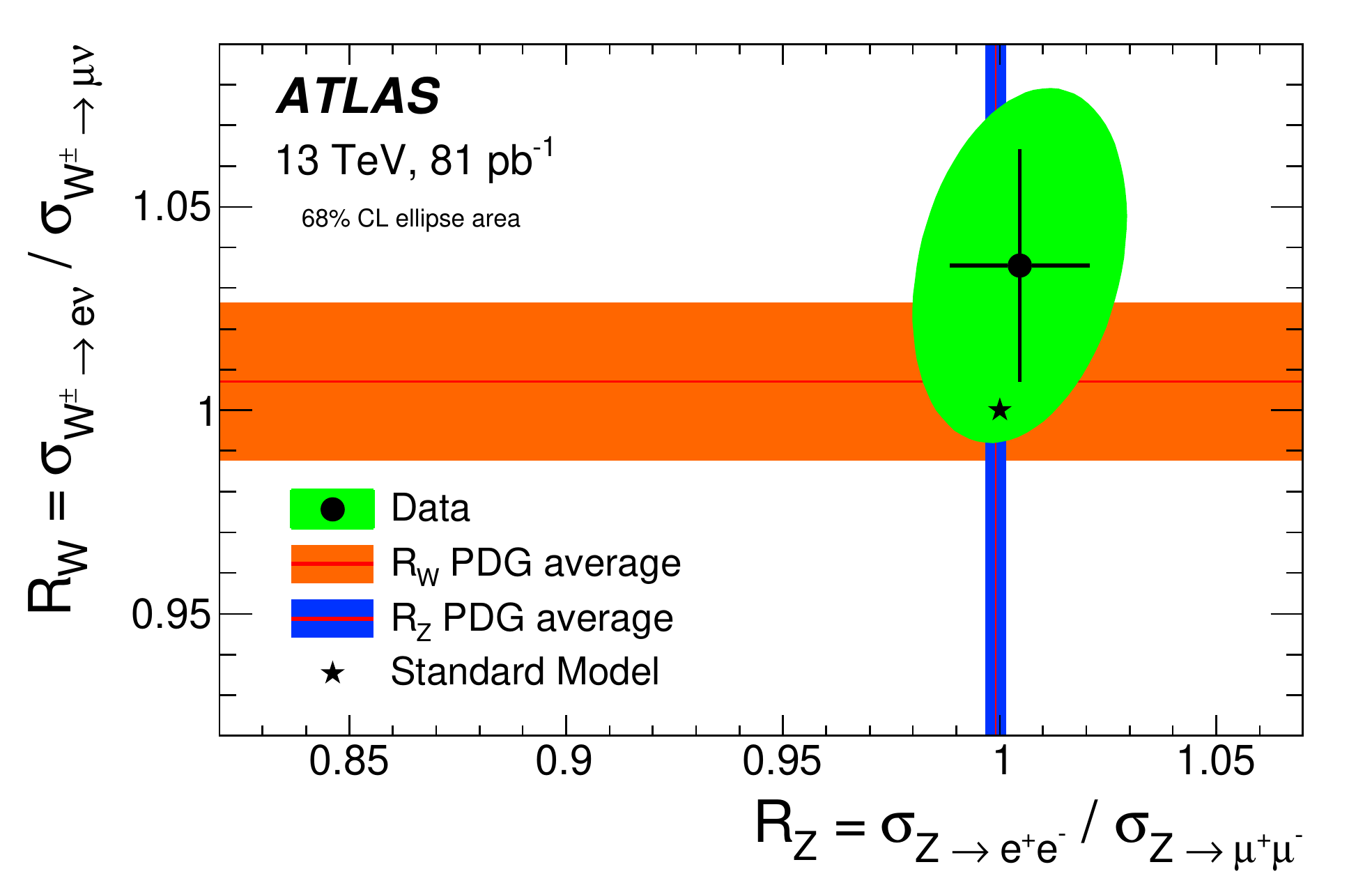} 
\caption{\label{fig:uni} 
Ratio of the electron- and muon-channel \Wboson\ and \Zboson-boson production fiducial cross sections, compared to
the expected values of the Standard Model of $(1,1)$ (neglecting mass effects that contribute at a level below $10^{-5}$) and previous experimental verifications of lepton universality for on-shell \Wboson and \Zboson bosons, shown as PDG average bands~\cite{PDG14,LEP-2}. The PDG average values and the result are shown with total uncertainties. 
}
\end{figure} 

\begin{figure}[t]
\centering
{\includegraphics[width=0.95\textwidth]{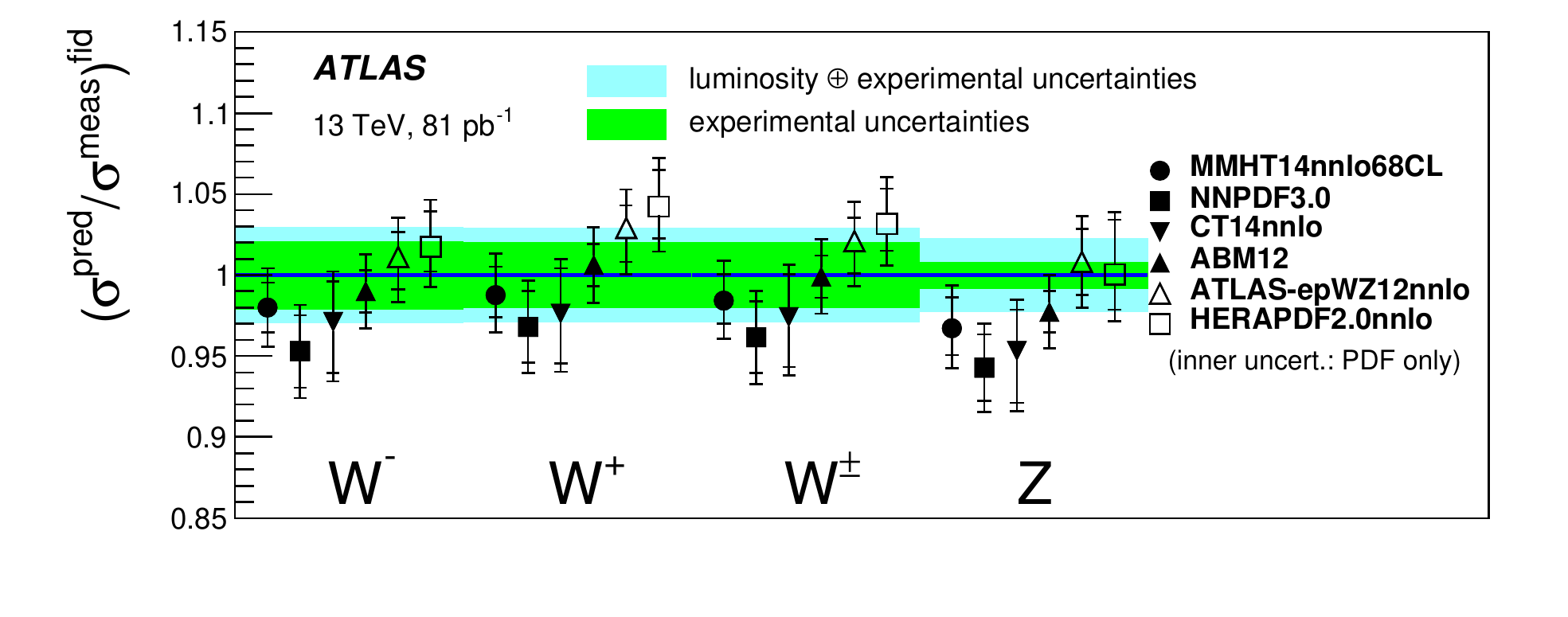}}

\caption{\label{fig:uni_ratios} 
Ratio of the predicted to measured fiducial cross section for the combined electron and muon channels using various PDFs.
The inner (outer) band corresponds to the experimental uncertainty without (with) the luminosity uncertainty. The inner error bar of the predictions represents the PDF uncertainty while the outer error bar includes the sum in quadrature of all other systematic uncertainties.
}
\end{figure}


The ratios of the fiducial electron and muon channel measurements in the \Wboson ($R_{\Wboson}=\sigma_{W^{\pm}\rightarrow e\nu}^{\rm{fid}}/\sigma_{W^{\pm}\rightarrow \mu\nu}^{\rm{fid}}$) and the \Zboson-boson ($R_{\Zboson}=\sigma_{Z\rightarrow e^+e^-}^{\rm{fid}}/\sigma_{Z\rightarrow \mu^+\mu^-}^{\rm{fid}}$) channels, 
evaluated taking into account correlations in the systematic uncertainties, are shown in Figure~\ref{fig:uni}.
Since these  results agree well with Standard Model expectations of lepton universality, a simultaneous combination of the
\Wplus, \Wminus and \Zboson-boson fiducial cross sections using the {\tt HERAverager} program~\cite{Glazov:2005rn,Aaron:2009bp} is performed.

The combination uses the individual sources of the systematic uncertainties, as shown in Table~\ref{tab:systotWZ}, in addition to uncertainties in the background evaluations. 
Sources corresponding to lepton reconstruction and identification are uncorrelated between the electron and muon channels.
Some sources, such as JES, JER, \MET and multijet background, only affect \Wboson-boson measurements.
The correlation model used for combining the multijet $W^+$ and $W^-$ uncertainties in each lepton channel is defined by: 
\begin{equation}
\delta(W^{\pm})^2 = \delta(W^+)^2 + \delta(W^-)^2 + 2\rho\delta(W^+)\delta(W^-) \mathrm{,} 
\label{eq:corrMod}
\end{equation}
performed separately for each source of systematic uncertainty considered for this background. All such uncertainties are considered to be uncorrelated between the electron and muon channels except that of the jet-energy-scale variation.
The correlation coefficient, $\rho$, is obtained from the uncertainties evaluated separately for the $W^+$ and $W^-$ channels ($\delta(W^+),\ \delta(W^-)$), and repeating the multijet background extraction without selecting the final-state charge ($\delta(W^{\pm})$).
The correlations of the systematic uncertainties vary from 0.2 to 1 (fully correlated), depending on the lepton channel and type of uncertainty.
The common normalisation uncertainty due to the luminosity calibration is excluded from the combination procedure and applied separately to the result. 

The combination yields a $\chi^{2}/N_\text{d.o.f.} = \AveChisq /3$, indicating excellent compatibility of the measurements.
Table~\ref{tab:pred_meas_xsections} gives the resulting combined cross sections. There is a reduction of uncertainty
compared to individual
electron and muon channel measurements since many of the systematic uncertainty sources are uncorrelated. The combined fiducial cross sections are extrapolated to the full phase space using the acceptance factors of Table~\ref{FiducialWZXSections}. These total cross sections are reported in Table~\ref{tab:pred_meas_xsections}.
The central values of the fiducial and total cross-section predictions, as described in Section~\ref{sec:method}, are also provided in Table~\ref{tab:pred_meas_xsections}. The  statistical uncertainties resulting from the evaluation of these predictions are negligible.

\begin{table}[t]
\begin{center}
\scriptsize
\begin{tabular}{c||c|c|c|c}
\hline
\hline
          & \multicolumn{2}{c|}{\bf Measured cross section $\times$ BR(\Wnocharge~$\rightarrow \ell\nu$, \Zboson~$\rightarrow \ell \ell$) [nb]} & \multicolumn{2}{c}{\bf Predicted cross section $\times$ BR(\Wnocharge~$\rightarrow \ell\nu$, \Zboson~$\rightarrow \ell\ell$) [nb]}  \\
         & \multicolumn{2}{c|}{(value $\pm$ stat $\pm$  syst $\pm$ lumi)}  & \multicolumn{2}{c}{(value $\pm$ PDF $\pm$  scale $\pm$ other)}  \\  
 Channel &  Fiducial & Total &  Fiducial & Total \\*[0.2cm] \hline \hline
\raisebox{-0.4ex}{\Wminus}  & \raisebox{-0.4ex}{$\AveCentralWm \pm \AveStatWm \pm \AveSystWm \pm \AveLumiWm $} & \raisebox{-0.4ex}{$\AveTotCentralWm \pm \AveTotStatWm \pm \AveTotSystWm \pm \AveTotLumiWm $} & \raisebox{-0.4ex}{$3.40^{+0.09}_{-0.11} \pm 0.04 \pm 0.06$}  & \raisebox{-0.4ex}{$8.54^{+0.21}_{-0.24} \pm 0.11 \pm 0.12$}  \\*[0.2cm]
\raisebox{-0.4ex}{\Wplus}  & \raisebox{-0.4ex}{$\AveCentralWp \pm \AveStatWp \pm \AveSystWp \pm \AveLumiWp $} & \raisebox{-0.4ex}{$\AveTotCentralWp \pm \AveTotStatWp \pm \AveTotSystWp \pm \AveTotLumiWp $}&  \raisebox{-0.4ex}{$4.42^{+0.13}_{-0.14} \pm 0.05 \pm 0.08$} & \raisebox{-0.4ex}{$11.54^{+0.32}_{-0.31} \pm 0.15 \pm 0.16$} \\*[0.2cm]
\raisebox{-0.4ex}{\Wpm}    & \raisebox{-0.4ex}{$\AveFidCentralW \pm \AveFidStatW \pm \AveFidSystW \pm \AveFidLumiW $} & \raisebox{-0.4ex}{$\AveTotCentralW \pm \AveTotStatW \pm \AveTotSystW \pm \AveTotLumiW $} &  \raisebox{-0.4ex}{$7.82^{+0.21}_{-0.25} \pm  0.09 \pm 0.13$} &  \raisebox{-0.4ex}{$20.08^{+0.53}_{-0.54} \pm 0.26 \pm 0.28$} \\*[0.2cm]
\hline
\raisebox{-0.4ex}{\Zboson} & \raisebox{-0.4ex}{$\AveCentralZ \pm \AveStatZ \pm \AveSystZ \pm \AveLumiZ$} & \raisebox{-0.4ex}{$\AveTotCentralZ \pm \AveTotStatZ \pm \AveTotSystZ \pm \AveTotLumiZ $} & \raisebox{-0.4ex}{$0.74^{+0.02}_{-0.03} \pm  0.01 \pm 0.01$} &  \raisebox{-0.4ex}{$1.89 \pm {0.05} \pm 0.03 \pm 0.03$} \\*[0.2cm] \hline \hline
         & \multicolumn{2}{c}{{\bf Measured ratio} (value $\pm$ stat $\pm$  syst)} & \multicolumn{2}{c|}{{\bf Predicted ratio} (value $\pm$ PDF)}  \\ \hline \hline
\raisebox{-0.4ex}{$\Wplus/\Wminus$}     &  \raisebox{-0.4ex}{$\WpWmcombVal \pm \WpWmcombSt \pm \WpWmcombSy$}                     &        \raisebox{-0.4ex}{--}         &            \raisebox{-0.4ex}{$1.30 \pm {0.01}$}               &  \raisebox{-0.4ex}{--}     \\*[0.2cm] 
\raisebox{-0.4ex}{$\Wboson/\Zboson$}   &  \raisebox{-0.4ex}{$ \WZcombVal \pm \WZcombSt \pm \WZcombSy$}                         &        --    &            \raisebox{-0.4ex}{$10.54 \pm {0.12}$ }             &  \raisebox{-0.4ex}{--}      \\*[0.2cm]                                                     
\hline \hline
\end{tabular}
\end{center}
\caption{The measured fiducial \sfid\ and total  \stot\ cross sections  for the combined electron and muon channels of \Wminus, \Wplus, \Wboson, and \Zboson-boson production and the fiducial ratios $\Wplus/\Wminus$ and $\Wboson/\Zboson$. Also shown are the predicted values as discussed in Section~\ref{sec:method}. The {\sc CT14}nnlo PDF is used for the predictions. In the second set of numerical columns, the error labelled as ``other'' represents the uncertainty in $\alpha_{\rm S}$ and in the beam energy. \label{tab:pred_meas_xsections} }
\end{table}

The combined fiducial cross sections are compared in Figure~\ref{fig:uni_ratios} to the predictions, which are 
calculated using different PDF sets. The measurements agree well with
the predictions and the experimental precision is comparable to the PDF uncertainties.

\begin{figure}[t]
  \begin{center}
\includegraphics[width=0.49\textwidth]{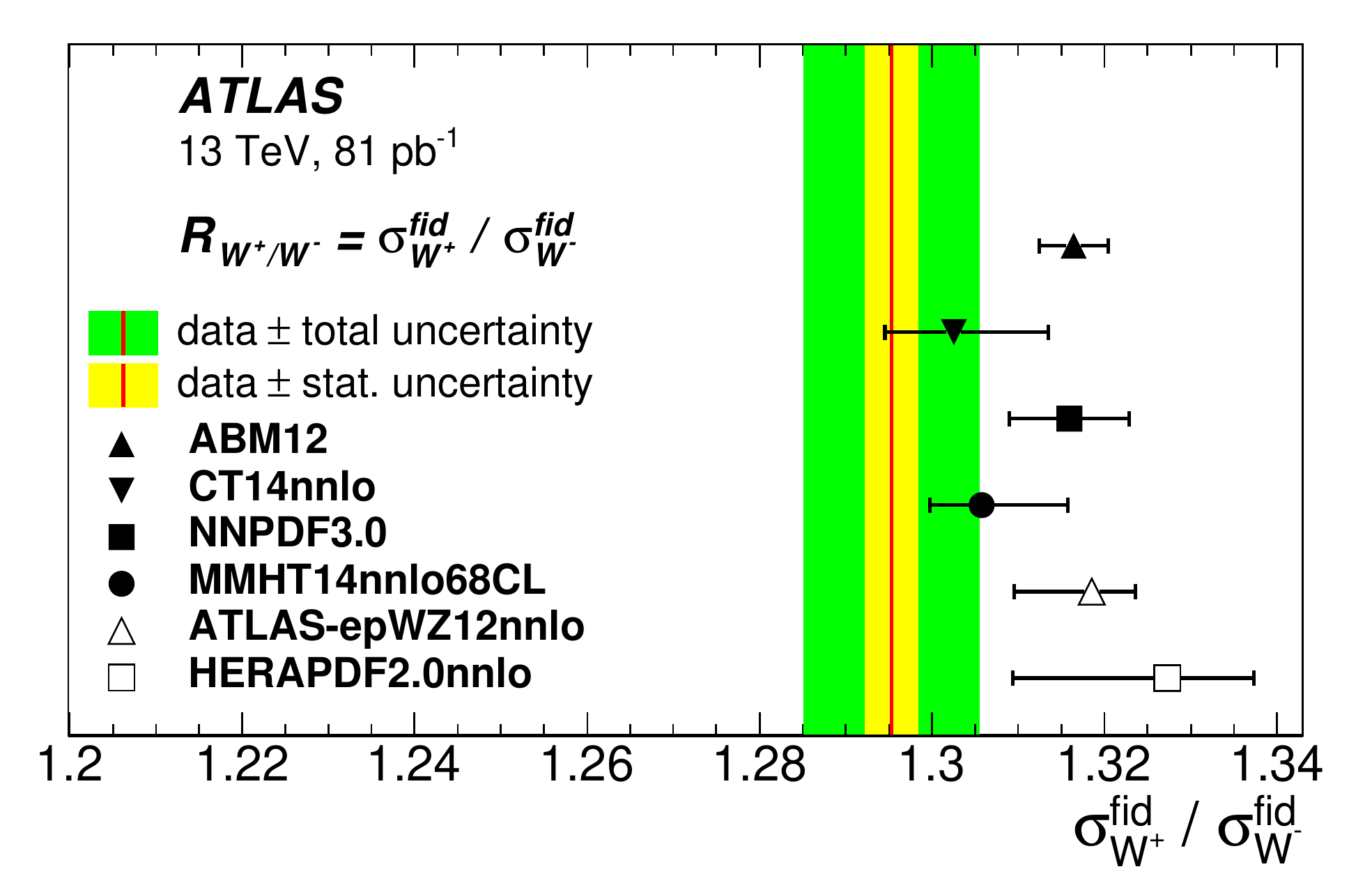}  
\includegraphics[width=0.49\textwidth]{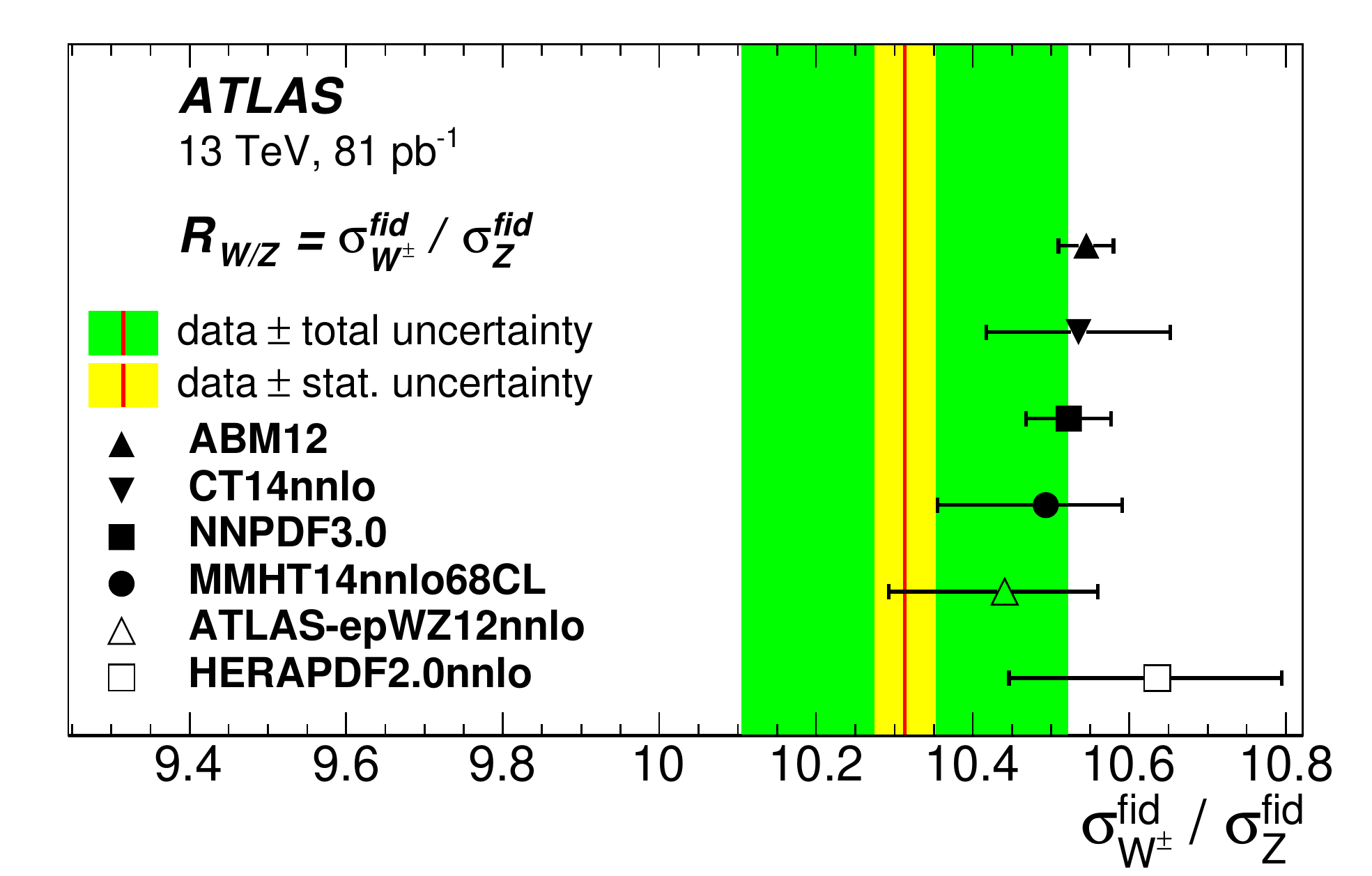}
 \caption{
Ratios (red line) of \Wplus to \Wminus boson (left) and \Wpm to \Zboson boson (right) combined production cross sections in the fiducial region compared to predictions based on different PDF sets.
The inner (yellow) shaded band corresponds to the statistical uncertainty while the outer (green) band shows statistical and
systematic uncertainties added in quadrature. The theory predictions are given with only the corresponding PDF
uncertainties shown as error bars. 
\label{fig:WZratio_predictions}}
\end{center}
\end{figure}

Ratios of the measured cross sections benefit from the cancellation of some experimental uncertainties.
The ratios of \Wplus to \Wminus and \Wpm to \Zboson-boson production, measured by the ATLAS, CMS, and LHCb
collaborations in the past~\cite{Aad:2011dm,CMS:2011aa,Chatrchyan:2014mua,Aaij:2015gna,Aaij:2015zlq}, proved to be powerful
tools to constrain PDF fits.  The ratio of  \Wplus to \Wminus-boson cross sections is mostly sensitive
to the difference of $u_{\rm v}$ and $d_{\rm v}$ valence-quark distributions at low Bjorken-$x$, while the ratio of \Wpm to \Zboson constrains the strange-quark distribution~\cite{Aad:2012sb}.  

The systematic uncertainties of the ratio measurements are largely uncorrelated between the electron and muon channels, apart 
from the common luminosity uncertainty.
However,  there is a  strong correlation between \Wplus and \Wminus-boson measurements and between the \Wpm
and \Zboson-boson results for the same-flavour measurement. 
The results for the measured \Wplus/\Wminus and \Wboson/\Zboson ratios of fiducial production cross sections in the combined electron and muon channels as well as the corresponding predictions as described in Section~\ref{sec:method} are given 
in Table~\ref{tab:pred_meas_xsections} and presented in Figure~\ref{fig:WZratio_predictions}. The dominant components of the systematic uncertainty in the \Wboson/\Zboson ratio are from both the multijet background and the jet-energy scale/resolution while that of the \Wplus/\Wminus ratio is from the uncorrelated part of the multijet background uncertainty.  
For the ratios $R_{\Wplus/\Wminus} = \sfidWp/\sfidWm$ and $R_{W/Z} =\sfidWpm/\sfidZ$, several predictions agree within quoted uncertainties, although all predictions are above the central value for the data in both cases.


\section{Conclusion}

Measurements with the ATLAS detector at the LHC of the \Wln\ and \Ztoll\ production cross sections based
on \ntotW\ and \ntotZ\ candidates, respectively, are presented. 
These results correspond to a total integrated luminosity of approximately \lumWZ of  proton--proton collisions at $\sqrt{s}=13$~TeV, the highest centre-of-mass energy ever available from a collider. The size of the \Wboson and \Zboson-boson production cross sections at this LHC Run-2 centre-of-mass energy are enhanced by a factor of nearly two from those at $\sqrt{s}=7$~TeV and 8~TeV in Run-1. 
The measurements of the fiducial cross sections of  \Wplus, \Wminus, and \Zboson-boson production are made separately in the electron and muon decay channels and are found to be consistent between the two channels. The datasets for electron and muon decay channels are then combined using a methodology which accounts for the correlations of the experimental systematic uncertainties.
The measured fiducial and total cross sections are found to agree with theoretical calculations based on NNLO QCD with NLO EW corrections. These measured cross sections have a global luminosity uncertainty of 2.1\%, while their remaining experimental uncertainties in the \Wboson and \Zboson-boson channels are found to be just under 3\% and 1\%, respectively.
The measurements of cross-section ratios benefit from the cancellation of some experimental
uncertainties, and are powerful tools to constrain PDF
fits. In particular, the fiducial cross-section ratio of \Wplus\
to \Wminus, measured with an uncertainty of 0.8\%, is able to discriminate amongst the  various PDF predictions presented. These results form a basis for further tests of perturbative QCD and exploration of the partonic content of the proton.

\section*{Acknowledgements}


We thank CERN for the very successful operation of the LHC, as well as the
support staff from our institutions without whom ATLAS could not be
operated efficiently.

We acknowledge the support of ANPCyT, Argentina; YerPhI, Armenia; ARC, Australia; BMWFW and FWF, Austria; ANAS, Azerbaijan; SSTC, Belarus; CNPq and FAPESP, Brazil; NSERC, NRC and CFI, Canada; CERN; CONICYT, Chile; CAS, MOST and NSFC, China; COLCIENCIAS, Colombia; MSMT CR, MPO CR and VSC CR, Czech Republic; DNRF and DNSRC, Denmark; IN2P3-CNRS, CEA-DSM/IRFU, France; GNSF, Georgia; BMBF, HGF, and MPG, Germany; GSRT, Greece; RGC, Hong Kong SAR, China; ISF, I-CORE and Benoziyo Center, Israel; INFN, Italy; MEXT and JSPS, Japan; CNRST, Morocco; FOM and NWO, Netherlands; RCN, Norway; MNiSW and NCN, Poland; FCT, Portugal; MNE/IFA, Romania; MES of Russia and NRC KI, Russian Federation; JINR; MESTD, Serbia; MSSR, Slovakia; ARRS and MIZ\v{S}, Slovenia; DST/NRF, South Africa; MINECO, Spain; SRC and Wallenberg Foundation, Sweden; SERI, SNSF and Cantons of Bern and Geneva, Switzerland; MOST, Taiwan; TAEK, Turkey; STFC, United Kingdom; DOE and NSF, United States of America. In addition, individual groups and members have received support from BCKDF, the Canada Council, CANARIE, CRC, Compute Canada, FQRNT, and the Ontario Innovation Trust, Canada; EPLANET, ERC, FP7, Horizon 2020 and Marie Sk{\l}odowska-Curie Actions, European Union; Investissements d'Avenir Labex and Idex, ANR, R{\'e}gion Auvergne and Fondation Partager le Savoir, France; DFG and AvH Foundation, Germany; Herakleitos, Thales and Aristeia programmes co-financed by EU-ESF and the Greek NSRF; BSF, GIF and Minerva, Israel; BRF, Norway; Generalitat de Catalunya, Generalitat Valenciana, Spain; the Royal Society and Leverhulme Trust, United Kingdom.

The crucial computing support from all WLCG partners is acknowledged
gratefully, in particular from CERN and the ATLAS Tier-1 facilities at
TRIUMF (Canada), NDGF (Denmark, Norway, Sweden), CC-IN2P3 (France),
KIT/GridKA (Germany), INFN-CNAF (Italy), NL-T1 (Netherlands), PIC (Spain),
ASGC (Taiwan), RAL (UK) and BNL (USA) and in the Tier-2 facilities
worldwide.

\clearpage

\printbibliography



\newpage 
\begin{flushleft}
{\Large The ATLAS Collaboration}

\bigskip

G.~Aad$^\textrm{\scriptsize 87}$,
B.~Abbott$^\textrm{\scriptsize 114}$,
J.~Abdallah$^\textrm{\scriptsize 65}$,
O.~Abdinov$^\textrm{\scriptsize 12}$,
B.~Abeloos$^\textrm{\scriptsize 118}$,
R.~Aben$^\textrm{\scriptsize 108}$,
O.S.~AbouZeid$^\textrm{\scriptsize 138}$,
N.L.~Abraham$^\textrm{\scriptsize 150}$,
H.~Abramowicz$^\textrm{\scriptsize 154}$,
H.~Abreu$^\textrm{\scriptsize 153}$
% ...
\bigskip
\\
% ...
$^{12}$ Institute of Physics, Azerbaijan Academy of Sciences, Baku, Azerbaijan\\
% ...
$^{65}$ University of Iowa, Iowa City IA, United States of America\\
% ...
$^{87}$ CPPM, Aix-Marseille Universit{\'e} and CNRS/IN2P3, Marseille, France\\
% ...
$^{108}$ Nikhef National Institute for Subatomic Physics and University of Amsterdam, Amsterdam, Netherlands\\
% ...
$^{114}$ Homer L. Dodge Department of Physics and Astronomy, University of Oklahoma, Norman OK, United States of America\\
% ...
$^{118}$ LAL, Univ. Paris-Sud, CNRS/IN2P3, Universit{\'e} Paris-Saclay, Orsay, France\\
% ...
$^{138}$ Santa Cruz Institute for Particle Physics, University of California Santa Cruz, Santa Cruz CA, United States of America\\
% ...
$^{150}$ Department of Physics and Astronomy, University of Sussex, Brighton, United Kingdom\\
% ...
$^{153}$ Department of Physics, Technion: Israel Institute of Technology, Haifa, Israel\\
$^{154}$ Raymond and Beverly Sackler School of Physics and Astronomy, Tel Aviv University, Tel Aviv, Israel\\
% ...
\end{flushleft}


\end{document}